\documentclass[fleqn,10pt]{wlscirep}
\usepackage[utf8]{inputenc}
\usepackage[T1]{fontenc}
\usepackage[justification=justified]{caption}
\usepackage[right]{lineno}

\title{Reproducible biomarkers: Leveraging nonlinear descriptors in the face of non-ergodicity}

\author[1,*]{Madhur Mangalam}
\author[2,3]{Arash Sadri}
\author[4]{Junichiro Hayano}
\author[5]{Eiichi Watanabe}
\author[6]{Ken Kiyono}
\author[7]{Damian G. Kelty-Stephen}
\affil[1]{\textit{Division of Biomechanics and Research Development, Department of Biomechanics, and Center for Research in Human Movement Variability, University of Nebraska at Omaha, Omaha, NE 68182, USA}}
\affil[2]{\textit{Lyceum Scientific Charity, Tehran, Iran}}
\affil[3]{\textit{Interdisciplinary Neuroscience Research Program, Students' Scientific Research Center, Tehran University of Medical Sciences, Tehran P94V+8MF, Iran}}
\affil[4]{\textit{Graduate School of Medicine, Nagoya City University, Nagoya, Aichi 467-8601, Japan}}
\affil[5]{\textit{Division of Cardiology, Department of Internal Medicine, Fujita Health University Bantane Hospital, Nagoya, Aichi 454-0012, Japan}}
\affil[6]{\textit{Graduate School of Engineering Science, Osaka University, Osaka 560-8531, Japan}}
\affil[7]{\textit{Department of Psychology, State University of New York at New Paltz, New Paltz, NY 12561, USA}}

\affil[*]{\href{mailto:mmangalam@unomaha.edu}{mmangalam@unomaha.edu}}

\begin{abstract}
Any reliable biomarker has to be specific, generalizable, and reproducible across individuals and contexts. The exact values of such a biomarker must represent similar health states in different individuals and at different times within the same individual to result in the minimum possible false-positive and false-negative rates. The application of standard cut-off points and risk scores across populations hinges upon the assumption of such generalizability. Such generalizability, in turn, hinges upon this condition that the phenomenon investigated by current statistical methods is ergodic, i.e., its statistical measures converge over individuals and time within the finite limit of observations. However, emerging evidence indicates that biological processes abound with non-ergodicity, threatening this generalizability. Here, we present a solution for how to make generalizable inferences by deriving ergodic descriptions of non-ergodic phenomena. For this aim, we proposed capturing the origin of ergodicity-breaking in many biological processes: cascade dynamics. To assess our hypotheses, we embraced the challenge of identifying reliable biomarkers for heart disease and stroke, which, despite being the leading cause of death worldwide and decades of research, lacks reliable biomarkers and risk stratification tools. We showed that raw R-R interval data and its common descriptors based on mean and variance are non-ergodic and non-specific. On the other hand, the cascade-dynamical descriptors, the Hurst exponent encoding linear temporal correlations, and multifractal nonlinearity encoding nonlinear interactions across scales described the non-ergodic heart rate variability ergodically and were specific. This study inaugurates applying the critical concept of ergodicity in discovering and applying digital biomarkers of health and disease.
\end{abstract}

\begin{document}

\flushbottom
\maketitle

\thispagestyle{empty}


\section*{Introduction}

Heart disease and stroke are the leading causes of disease and disability globally and in the United States, claiming $655,000$ American lives every year---one in four deaths \cite{tsao2023heart,roth2020global}. This staggering toll of cardiovascular diseases does not end here, as it costs the nation over $\$200$ billion annually in direct medical expenses and lost productivity. This colossal burden highlights the importance of early diagnosis and intervention of heart disease and stroke. One of the primary requisites for effective diagnosis is the availability of specific and reliable biomarkers. Although numerous biomarkers, risk stratification models, and risk scores for various cardiovascular diseases have been proposed over the past decades, effective diagnostic and prognostic digital biomarkers are still missing \cite{omland2017state,muanescu2022conventional,pal2023emerging}. The urgency of addressing this need is amplified by the rise and ever-growing expansion of diverse digital health and telehealth solutions in recent years, specifically in the cardiovascular field \cite{zwack2023evolution,yeung2022research,hughes2023wearable}. Such solutions, like mobile applications (mhealth), smart watches, wearable devices, implantable electronic devices, and implantable hemodynamic monitors, enable the gathering of vast amounts of data for everyone; however, the lack of diagnostic and prognostic biomarkers lays waste to this ability as such valuable amounts of data cannot be appropriately used. Lack of evidence of effect has been cited as one of the reasons why digital health technologies have not been widely employed in clinical settings \cite{nes2020digital}. The lack of reliable digital biomarkers can be considered one of the main contributors to this lack of evidence. 

Heart rate variability (HRV) has been one of the key \textit{noninvasive biomarkers} of cardiovascular health \cite{karemaker2017introduction}. It measures the fluctuations and variations in time intervals between successive heartbeats or R-R intervals (RRi). HRV is an emergent phenomenon that emerges out of the complex and nonlinear interactions between the cardiovascular and nervous systems \cite{agliari2020detecting,freeman2006autonomic,zhong2007autonomic} and represents the peripheral output of the central autonomic network (CAN) and the capacity for behavioral adaption to environmental stresses \cite{beissner2013autonomic,brosschot2017exposed,cechetto2009functional,guyenet2006sympathetic,ruiz2016human,thayer2009claude,thayer2012meta,thayer2021stress,wulsin2018stress}. Because it emerges from such complex and integral interactions, HRV can be a representative marker of cardiovascular health. Healthy human HRV indicates desirable balance and interaction between the functions of the sympathetic and parasympathetic nervous systems \cite{bootsma1994heart,khan2019heart,van1993heart}. Group-level findings have shown that abnormal HRV indicates an imbalance between the two systems, among other pathophysiologies, and is associated with an increased risk of heart disease, including myocardial infarction, heart failure, and sudden cardiac death \cite{amezquita2018novel,carney2001depression,chattipakorn2007heart,fei1996short,hillebrand2013heart,la2003short,lombardi2001sudden,makikallio2001prediction,ponikowski1997depressed,sessa2018heart,woo1992patterns,xhyheri2012heart}. HRV is also a potential predictor of morbidity and mortality in other diverse disorders, including type $2$ diabetes \cite{benichou2018heart,cha2018time,gottsater2006decreased,kataoka2004low,may2011long}, chronic obstructive pulmonary disease (COPD) \cite{bedard2010reduced,carvalho2011fractal,gunduz2009heart,lacasse2005post,volterrani1994decreased}, chronic kidney disease (CKD) \cite{chandra2012predictors,drawz2013heart,fukuta2003prognostic,oikawa2009prognostic,ranpuria2008heart}, dementia \cite{da2018heart}, depressive disorders \cite{koch2019meta,schiweck2019heart}, anxiety and stress disorders \cite{chalmers2014anxiety,thayer2012meta}, obsessive compulsive disorder \cite{olbrich2022heart}, autism spectrum disorders \cite{cheng2020heart}, and attention-deficit/hyperactivity disorder \cite{robe2019attention}. Some studies have suggested that HRV might be superior to many other biomarkers in representing the overall state of health and well-being \cite{jarczok2013heart,jarczok2015investigating}. These developments have sparked considerable optimism surrounding the potential of heart rate-sensing wearable devices to detect and track cardiovascular diseases \cite{avram2019real,bayoumy2021smart,holko2022wearable,lee2018highly,pereira2020photoplethysmography,owens2020role,raza2022intelligent}.

Although the emergence of HRV out of complex and intricate interactions confers HRV such an ability to represent the state of the body, it also makes its appropriate application as a digital biomarker replete with nuances. Analyses of heartbeat dynamics and HRV reveal significant nonlinearity, non-Gaussianity, and chaotic behaviors in RRi series \cite{captur2017fractal,gieraltowski2012multiscale,goldberger2002fractal,ivanov1996scaling,ivanov2001,kiyono2004critical,kiyono2005phase,kurths1995quantitative,lefebvre1993predictability,lin2001modeling,peng1995fractal,perkiomaki2011heart,sugihara1996nonlinear,tan2009fractal,voss2009methods}. These statistical signatures of nonlinearity, non-Gaussianity, and chaotic behaviors in RRi can be interpreted as manifestations of the emergence of HRV from interdependent and bidirectional interactions across multiple timescales. Such processes which lead to multiplicative fluctuations and dynamics have been termed \textit{multifractal cascades} \cite{kelty2013tutorial,turcotte2002self,lovejoy2006multifractals,olsson2007analysis,ihlen2010interaction}. The cascade dynamical nature of HRV, like many other behavioral and physiological functions \cite{bloomfield2021perceiving,kelty2021multifractal,kelty2022turing,kelty2023multifractalnonlinearity,mangalam2020global,mangalam2020multifractal,mangalam2020multiplicative}, inclines many of its measurements and descriptors toward a characteristic that has been, unfortunately, grossly overlooked in the biomedical literature: \textit{ergodicity}. We believe the overlooking of ergodicity has hindered the broad application of HRV probably much more than the other challenges which have been discussed regarding HRV, like analytical challenges associated with data variability, missing data and artifacts, and lack of theory for data interpretation \cite{biswas2019heart,cajal2022effects,georgiou2018can,isakadze2020useful,karemaker2020interpretation,karemaker2022multibranched,nelson2019accuracy,raja2019apple,shaffer2014healthy,hayano2019pitfalls,shaffer2017overview,thomas2019validity}.

\begin{figure}[bt!]
\centering
\includegraphics[width=5.875in]{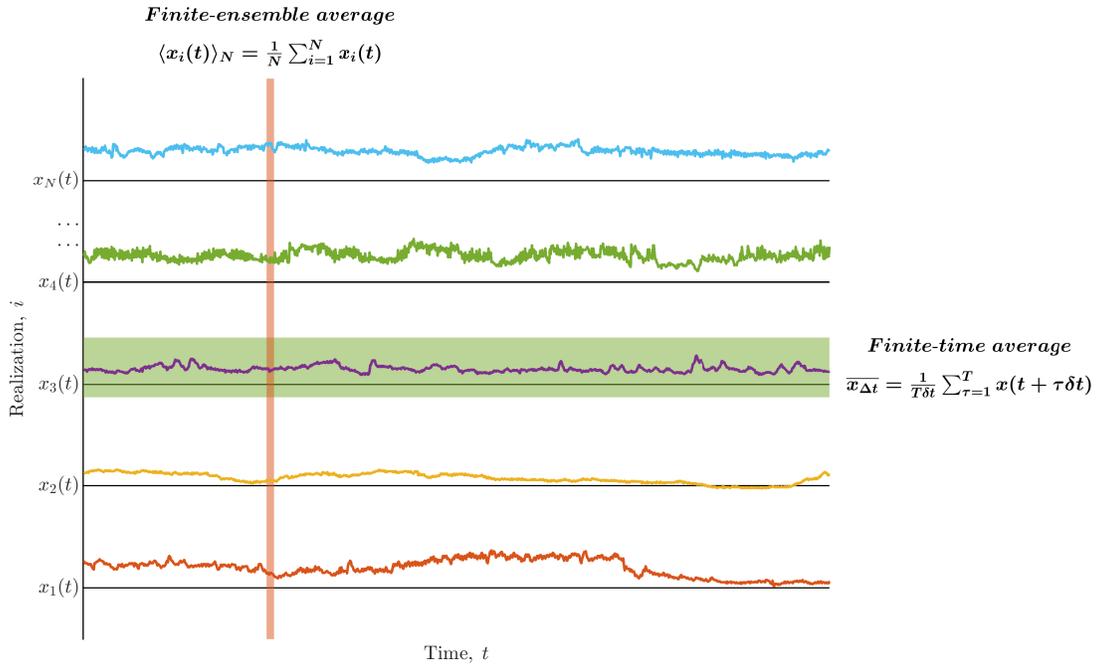}
\caption{\textbf{Non-ergodicity refers to the lack of equivalence between finite-ensemble and finite-time averages.} The finite-ensemble average, which biomedical discourse recognizes as the ``sample average,'' is $\langle x_{i}(t) \rangle _{N} = \frac{1}{N} \sum_{i=1}^{N} x_{i}(t)$, where $x_{i}(t)$ is the $i$th of $N$ individual cases of $x(t)$ included in the finite-ensemble average. The finite-time average when the measured behavior $x$ changes at $T = \Delta t/\delta t$ discrete times $t + \delta t, t + 2 \delta t, \dots$ is $\overline{x_{\Delta t}} = \frac{1}{T \delta t} \sum_{\tau = 1}^{T} x(t + \tau \delta t)$.}
\label{fig: Ergodicity}
\end{figure}

Ergodicity is an essential requirement of a digital biomarker to be applied reliably in current medical practice. Similar values of a digital biomarker across different individuals must represent similar bodily states. In other words, standard cut-off points of such a biomarker must reliably separate the states of health and disease in each different individual. The importance of this requirement is highlighted when we pay attention to the research designs and statistical practices that have dominated scientific investigations for about a century. These research designs and statistical practices were devised in the twentieth century by pioneers like Francis Galton \cite{galton1883inquiries,galton1889natural}, Karl Pearson \cite{neyman1933ix}, Ronald Fisher \cite{fisher1922mathematical,fisher1949design,fisher1950statistical}, Jerzy Neyman \cite{neyman1933ix}, Egon Pearson\cite{neyman1933ix}, and Udny Yule \cite{aldrich2016origins}. Based on these practices, most medical research, similar to most biological, psychological, and social research, has been aggregating the data gathered from randomly selected groups of individuals and used group-based statistical methods to reach conclusions. Such conclusions are then deemed generalizable to the behaviors of different individuals across different contexts. However, ergodicity is a requisite of this generalizability from group-level data to an individual's behaviors. In non-ergodic measurements, the behaviors of an individual at a specific time diverge from the average of that measurement across a group of individuals and also the average of that individual's behaviors over an extended period \cite{molenaar2004manifesto,molenaar2009new,voelkle2014toward,mangalam2021point}. Ergodicity refers to the convergence of these two averages: the finite-ensemble average and the finite-time average (\textbf{Fig.~\ref{fig: Ergodicity}}). The finite-ensemble average, which is also recognized as the ``sample average,'' is
\begin{equation*}
  \langle x_{i}(t) \rangle _{N} = \frac{1}{N} \sum_{i=1}^{N} x_{i}(t), \tag{1}\label{eq:1}
\end{equation*}
where $x_{i}(t)$ is the $i$th of $N$ individual cases of $x(t)$ included in the finite-ensemble average. The finite-time average, which biomedical discourse recognizes as the ``average performance/trajectory of the individual,'' is
\begin{equation*}
  \overline{x_{\Delta t}} = \frac{1}{\Delta t} \int_{t}^{t + \Delta t} x(t)dt, \tag{2}\label{eq:2}
\end{equation*}
for continuous change. The finite-time average when the measured behavior $x$ changes at $T = \Delta t/\delta t$ discrete times $t + \delta t, t + 2 \delta t, \dots$ is
\begin{equation*}
  \overline{x_{\Delta t}} = \frac{1}{T \delta t} \sum_{\tau = 1}^{T} x(t + \tau \delta t). \tag{3}\label{eq:3}
\end{equation*}
So, ergodicity is an equivalence between these two averages,
\begin{equation*}
  \lim_{\Delta t \to \infty} \frac{1}{\Delta t} \int_{t}^{t + \Delta t} x(t)dt = \lim_{N \to \infty} \frac{1}{N} \sum_{i=1}^{N} x_{i}(t). \tag{4}\label{eq:4}
\end{equation*}

Another phrasing of the concept of ergodicity is that ergodic systems visit all of their possible states---in a sense, ergodic systems do not have a deep sense of ``history.’’ The criterion of ``mixing’’ emphasizes this addition to the traditional interpretation of ergodicity. Mixing denotes independence of the states of a system across time in a way that all values of a stochastic process across all times would have equal probabilities \cite{kelty2022fractal}. This concept clarifies why emerging experimental data suggests that the processes related to organisms teem with, and probably are even dominated by, non-ergodicity \cite{fisher2018lack,li2022non,weigel2011ergodic}, although the inferences of the majority of biological, psychological, and social studies in the past century have been based on this implicit presupposition that the processes they study and their measurements are ergodic. Biological processes teem with properties like interactions across space and time scales \cite{ihlen2010interaction,kelty2013tutorial}, historical contingency \cite{card2019historical,xie2021contingency}, and context dependency to break ergodicity \cite{voelkle2014toward}. Consider the exemplary biological process we have chosen in this study: HRV. As we mentioned earlier, data strongly suggests that heartbeat dynamics and HRV have a cascade dynamical nature and emerge from interdependent and bidirectional interactions across scales \cite{captur2017fractal,gieraltowski2012multiscale,goldberger2002fractal,ivanov1996scaling,ivanov2001,kiyono2004critical,kiyono2005phase,kurths1995quantitative,lefebvre1993predictability,lin2001modeling,peng1995fractal,perkiomaki2011heart,sugihara1996nonlinear,tan2009fractal,voss2009methods}. Also, HRV and many of its descriptors highly depend on various individual, contextual, and measurement factors such as sex \cite{koenig2016sex}, age \cite{abhishekh2013influence,almeida2016aging,bonnemeier2003circadian}, aerobic fitness and physical activity \cite{aubert2003heart,de1993heart,kiyono2005phase}, smoking \cite{bodin2017association,uusitalo2007role}, consumption of coffee \cite{uusitalo2007role} and alcohol \cite{schnell2013effects}, medications \cite{uusitalo2007role}, environmental noise and CO concentrations \cite{schnell2013effects}, respiration \cite{hirsch1981respiratory,shaffer2014healthy}, pace of breathing \cite{nunan2010quantitative}, times of sleep \cite{faust2020deviations}, posture \cite{da2019impact,porta2023changes}, the length of the observation period \cite{laborde2017heart}, the time of measurement \cite{kiyono2005phase,lord2001low}, the used detection method \cite{jeyhani2015comparison}, sampling frequency \cite{merri1990sampling}, and procedures of processing and removing artifacts \cite{bassi2018inter,farah2016intra,peltola2012role,plaza2020inter}. Such historical contingency and context-dependency of HRV and other biological processes generally lead to non-ergodicity and lack of generalizability from group-level findings to individuals \cite{mangalam2021point,voelkle2014toward}.

The concern for ergodicity is evident in the application of HRV. An appropriate diagnosis and risk stratification based on HRV depends on two conditions: First, the limited data gathered during the visits, consultations, or laboratory assays should sufficiently represent the states of the individual's body over time. Second, the standard and established principles and cut-off points used to make decisions should be generalizable to that individual. These conditions have been taken for granted until now. However, as we discussed, evidence suggests that these conditions are probably violated because of non-ergodicity. Neglecting this violation can be detrimental; for instance, if screening is conducted on the entire general population, a minor increase in false positive rate can hugely raise subsequent medical tests and expenses \cite{bayoumy2021smart,isakadze2020useful,marcus2020apple}. An increased false-negative rate also implies delayed anticoagulant medication and increased risk of stroke in symptomatic or high-risk patients.

Our concern for ergodicity is not restricted to the application of HRV. We \cite{kelty2022fractal,kelty2023multifractal,mangalam2021point,mangalam2022ergodic}, alongside a few others \cite{fisher2018lack,molenaar2004manifesto,molenaar2009new,voelkle2014toward}, believe that ergodicity is an integral concept that undermines how scientific research across diverse fields has tried to identify cause-effect relationships. The breaking of ergodicity is abundant in biological processes and invalidates many conclusions of group-based research designs and statistical methods. Indeed, neglecting this non-ergodicity and lack of generalizability could be the leading cause of the reproducibility crisis \cite{fisher2018lack,mangalam2021point}, which currently encompasses diverse fields from biomedical and psychological sciences to social sciences and economics \cite{hanin2017statistical,ioannidis2005most}. Specifically, in applying HRV as a biomarker of health and disease, some studies have suggested that the irreproducibility of results could be a critical problem \cite{bassi2018inter,farah2016intra,da2019impact,hojgaard2005reproducibility,leicht2008moderate,lord2001low,pitzalis1996short,plaza2020inter,sacha2013heart,sandercock2005reliability}.

This study is an attempt in continuation of our previous works to obtain a solution to the problem of making generalizable inferences about non-ergodic processes. In this series of works, we first tried identifying sources of non-ergodicity in biological processes. Having recognized the abundance of multifractal and cascade dynamics in biological processes \cite{bloomfield2021perceiving,kelty2021multifractal,kelty2022turing,kelty2023multifractalnonlinearity,mangalam2020global,mangalam2020multifractal,mangalam2020multiplicative}, we hypothesized that a potential source of non-ergodicity could be the emergence of many biological processes out of interdependent and bidirectional interactions across spatial and temporal scales, as in cascades. We observed phenomena that corroborated this hypothesis \cite{kelty2022fractal,kelty2023multifractal,mangalam2022ergodic,mangalam2023ergodic}. Afterward, interestingly, we observed that descriptors that could capture the cascade-dynamical sources of ergodicity breaking in a process might provide ergodic descriptions of that process \cite{kelty2022fractal,kelty2023multifractal,mangalam2022ergodic,mangalam2023ergodic}.

Here, prompted by the huge amount of evidence that had suggested the multifractal and cascade-dynamical nature of HRV \cite{captur2017fractal,gieraltowski2012multiscale,goldberger2002fractal,ivanov1996scaling,ivanov2001,kiyono2004critical,kiyono2005phase,kurths1995quantitative,lefebvre1993predictability,lin2001modeling,peng1995fractal,perkiomaki2011heart,sugihara1996nonlinear,tan2009fractal,voss2009methods}, We hypothesized that this nature of HRV leads to the non-ergodicity of this phenomenon. Consequently, We predicted that the linear commonly used descriptors of HRV and raw RRi series, like sample means and variances, would be non-ergodic and lack generalizability and reproducibility. Afterward, we hypothesized that descriptors that would capture the source of the non-ergodicity of HRV might provide ergodic descriptions of this non-ergodic phenomenon. Descriptors of the nonlinear, non-Gaussian, multifractal, and cascade-dynamical behaviors of HRV, some of which we had developed in our previous works, seemed worthy candidates \cite{gieraltowski2012multiscale,kiyono2005multiscale,hayano2011increased,kiyono2004critical,kiyono2005multiscale,kiyono2005phase,kiyono2007estimator,kiyono2008non,kiyono2009log,kiyono2012non}. For this study, we chose descriptors of long-range correlations, $H_{fGn}$, and multifractal nonlinearity, $t_{MF}$. We found strong support for our hypotheses.

\section*{Results}

We analyzed the long-term ambulatory HRV in $108$ chronic heart failure (CHF) patients---$69$ survivors (age (\textit{mean}$\pm$\textit{SD}) = $64 \pm 15$ years; $27$ women) and $39$ nonsurvivors ($70 \pm 14$ years; $20$ women)---who died due to any cause within the follow-up period of $33\pm17$ months, and $115$ age-matched healthy older adults ($47.7 \pm 18.2$ years; $25$ women). The endpoint was all-cause mortality. The majority of deaths ($34/39$) were cardiac-related, including death from progressive heart failure ($n=23$), sudden death ($n=10$), and acute myocardial infarction ($n=1$). The remaining five patients died of sepsis ($n=1$), pneumonia ($n=3$), and stroke ($n=1$). We reanalyzed HRV data from one of our previous published studies \cite{kiyono2008non}. \textbf{Table~\ref{tab: Table1}} summarizes the demographic and baseline clinical characteristics of the CHF patients.

\begin{table}[bt!]
\centering
\begin{tabular}{|l|l|l|l|}
\hline
\textbf{Characteristics} & \textbf{Nonsurvivors} $(n=39)$ & \textbf{Survivors} $(n=69)$ \\
\hline
Age (years) & $70 \pm 14$ & $64 \pm 15$ \\
Sex (M/F) & $19/20$ & $42/27$ \\
\hline
New York Heart Association functional class & & \\
II & $3(8\%)$ & $3(13\%)$ \\
III-IV & $36(92\%)$ & $60(87\%)$ \\
Ischemia & $17(43\%)$ & $19(28\%)$ \\
Left ventricular ejection fraction (\%)	 & $40 \pm 12$ & $39 \pm 14$ \\
BNP (pg/mL) & $1,225 \pm 903$ & $704 \pm 606$  \\
 ln BNP & $6.8 \pm 0.8$ & $6.1 \pm 1.1$ \\
BUN (mg/dL) & $32 \pm 18$ & $23 \pm 13$ \\
 ln BUN & $3.3 \pm 0.5$ & $3.0 \pm 0.5$ \\
Cr (mg/dL) & $1.7 \pm 1.3$ & $1.1 \pm 1.0$ \\
 ln Cr & $0.25 \pm 0.69$ & $-0.13 \pm 0.60$ \\
 \hline
Medication at Holter recording & &\\
Beta-blocker & $11(28\%)$ & $23(33\%)$ \\
ACE/ARB & $19(49\%)$ & $32(46\%)$ \\
Loop diuretic & $26(67\%)$ & $30(43\%)$ \\
Spironolactone & $16(41\%)$ & $17(25\%)$ \\
\hline
Medication before hospital discharge & & \\
Beta-blocker & $26(67\%)$ & $48(70\%)$ \\
ACE/ARB & $26(67\%)$ & $55(80\%)$ \\
Loop diuretic & $27(95\%)$ & $62(90\%)$ \\
Spironolactone & $23(59\%)$ & $40(58\%)$ \\
Ventricular premature beats per hour & $22 \pm 64$ & $24 \pm 70$ \\
\hline
\end{tabular}
\newline
BNP = brain natriuretic protein; BUN = blood urea nitrogen; Cr = creatinine; ACE = angiotensin-converting enzyme inhibitor; ARB = angiotensin II receptor blocker.
\caption{\label{tab: Table1} \textbf{Baseline clinical characteristics of the chronic heart failure patients.} Reproduced from Kiyono et al. \cite{kiyono2008non}.}
\end{table}

\subsection*{HRV breaks ergodicity}

\begin{figure}[bt!]
\centering
\includegraphics[width=6.125in]{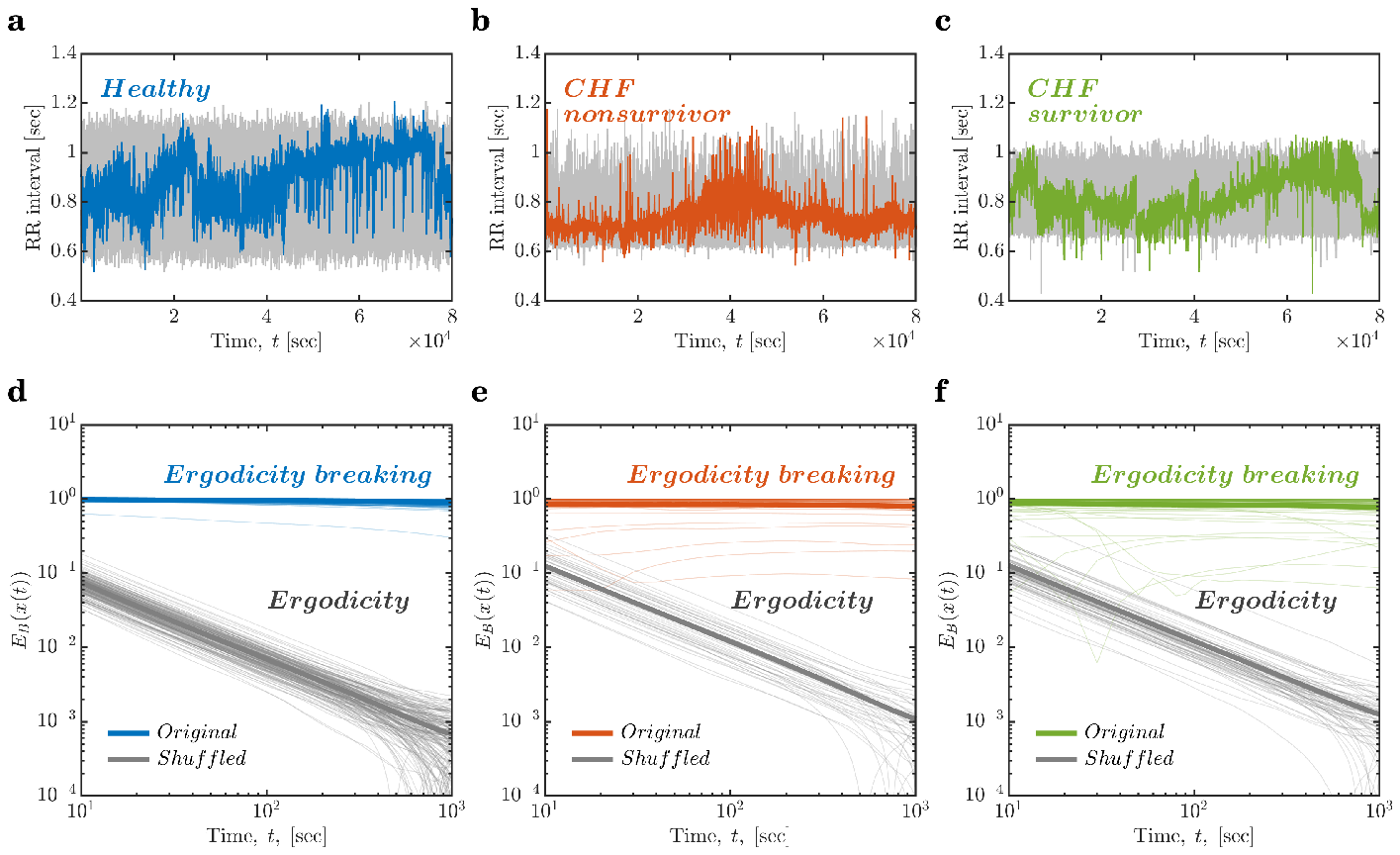}
\caption{\textbf{The raw R-R interval (RRi) series are non-ergodic.} \textbf{a--c.} Representative examples of the original and shuffled RRi series (\textit{colored lines} and \textit{grey lines}, respectively). \textbf{a.} RRi series for a healthy control (a $54$-year-old woman). \textbf{b.} RRi series for a $74$-year-old man with congestive heart failure (CHF) who died $101$ days after the measurement. \textbf{c.} RRi series for an $82$-year-old woman who survived CHF. The original RRi series, for healthy controls \textbf{d.} as well as the two patient groups \textbf{e, f.}, show no change in the ergodicity breaking parameter, $E_{B}$, over progressively longer periods, reflecting that HRV breaks ergodicity (\textit{colored lines}). Shuffling the original RRi series produces an RRi series that is ergodic, as indicated by the rapid decay in $E_{B}$ over progressively longer periods (\textit{grey lines}). \textit{Thin lines} and \textit{thick lines} in \textbf{d--f} represent ergodicity breaking for individuals and mean ergodicity breaking for the three groups, respectively.}
\label{fig: HRV-1}
\end{figure}

To examine the ergodic properties of the RRi series, we submitted the original RRi series and the corresponding shuffled versions to the Thirumalai-Mountain analysis \cite{he2008random,thirumalai1989ergodic}, which yields a dimensionless metric called the ergodicity breaking factor, $E_{B}$,
\begin{equation*}
  E_{B}(x(t)) = \frac{ \Bigl \langle \Bigl [ \overline{\delta^{2}(x(t))} \Bigl ]^{2} \Bigl \rangle - \Bigl \langle \overline{\delta^{2}(x(t))} \Bigl \rangle^{2}}{ \Bigl \langle \overline{\delta^{2}(x(t))} \Bigl \rangle ^{2}}. \tag{5}\label{eq:5}
\end{equation*}
where $\overline{\delta^{2}(x(t))} = \int_{0}^{t - \Delta} [x(t^{\prime} + \Delta) - x(t^{\prime})]^{2}dt^{\prime} \bigl / (t - \Delta)$ is the time average mean-squared displacement of the stochastic series $x(t)$ for lag time $\Delta$. Rapid decay of $E_{B}$ to a finite asymptotic value for progressively larger samples, i.e., $E_{B}\rightarrow0$ as $t \rightarrow \infty$ implies ergodicity. Slower decay indicates less ergodic systems in which trajectories are less reproducible. No decay or convergence to a finite asymptotic value indicates strong ergodicity breaking \cite{deng2009ergodic,wang2020fractional}. $E_{B}(x(t))$ thus allows testing whether a given series breaks ergodicity. $E_{B}$ for the original RRi series did not decay at all with $t$, essentially remaining unchanged over a progressively longer time for healthy controls as well as the two patient groups ($E_{B}(x(t)) = -0.0146\frac{\Delta}{t}, 0.0002\frac{\Delta}{t}$, and $-0.0182\frac{\Delta}{t}$ for healthy controls, CHF nonsurvivors, and CHF survivors, respectively; \textit{colored lines} in \textbf{Figs.~\ref{fig: HRV-1}d--f}). These values of $E_{B}(x(t))$ indicate strong ergodicity breaking in the original RRi series. In contrast, $E_{B}$ for the shuffled RRi series rapidly decayed to a finite asymptotic value, indicating ergodicity ($E_{B}(x(t)) = -1.0227\frac{\Delta}{t}, -1.0168\frac{\Delta}{t}$, and $-1.0145\frac{\Delta}{t}$ for healthy controls, CHF nonsurvivors, and CHF survivors, respectively; \textit{grey lines} in \textbf{Figs.~\ref{fig: HRV-1}d--f}). As by shuffling the original RRi series, the temporal structure and information of the RRi series are removed, these values of $E_{B}(x(t))$ suggest that the very temporal structure of HRV is the source of non-ergodicity in HRV.

\subsection*{Linear descriptors based on mean and variance are non-ergodic}

\begin{figure}[bt!]
\centering
\includegraphics[width=7in]{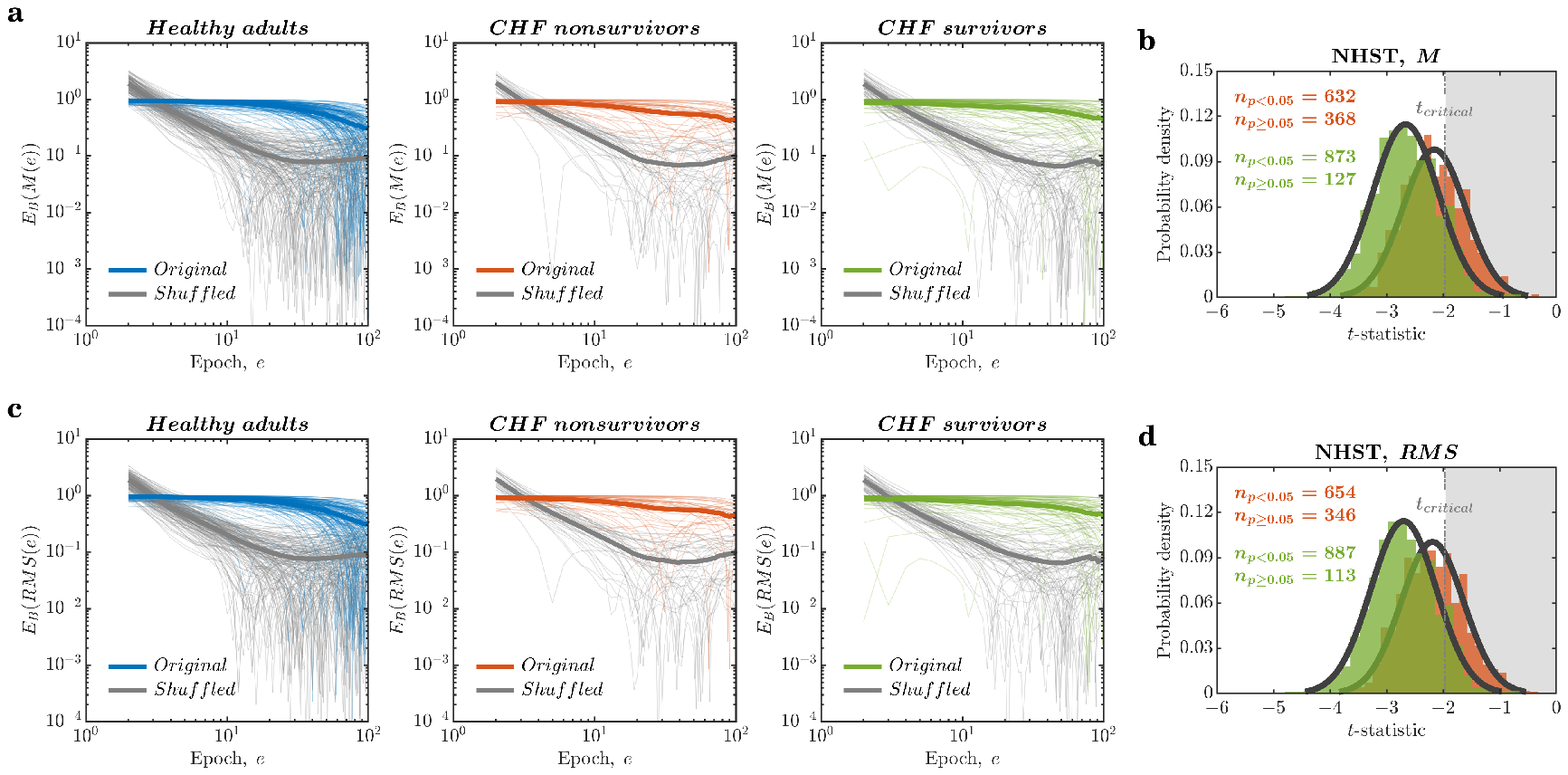}
\caption{\textbf{Commonly used linear descriptors of HRV based on mean and root mean square are non-ergodic.} \textbf{a.} The ergodicity breaking parameter, $E_{B}$, did not change for the mean of successive RR intervals, $M$, in the original RRi series over a progressively larger number of epochs (\textit{colored lines}). In contrast, $E_{B}$ decayed rapidly for the $M$ of the shuffled RRi series (\textit{grey lines}). \textbf{b.} Null hypothesis significance testing (NHST) for $M$ across the three groups. One-way ANOVAs failed to detect reduced $M$ of HRV due to the CHF, $36.8\%$ and $12.7\%$ times in nonsurvivors (\textit{red histogram}) and survivors (\textit{green histogram}), respectively, compared to healthy controls. \textbf{c.} $E_{B}$, did not change for the root mean square of successive RR interval differences, $RMS$, in the original RRi series over a progressively larger number of epochs (\textit{colored lines}). In contrast, $E_{B}$ decayed rapidly for the $RMS$ of the shuffled RRi series (\textit{grey lines}). \textbf{d.} Null hypothesis significance testing (NHST) for $RMS$ across the three groups. One-way ANOVAs failed to detect reduced $M$ of HRV due to the CHF, $34.6\%$ and $11.3\%$ times in nonsurvivors (\textit{red histogram}) and survivors (\textit{green histogram}), respectively, compared to healthy controls. \textit{Thin lines} and \textit{thick lines} in \textbf{a, c} represent $E_{B}$ for individuals and mean $E_{B}$ for the three groups, respectively.}
\label{fig: HRV-2}
\end{figure}

Now that we have witnessed ergodicity breaking in the raw RRi series, let us investigate the ergodic properties of some of the linear descriptors widely used in cardiovascular digital medicine: HRV parameters based on mean and variance \cite{shaffer2017overview}. Here, we chose the mean and root mean square of successive RR interval differences, hereinafter noted as $M$ and $RMS$, respectively. Similar to the behavior of $E_{B}$ for the raw RRi series, $E_{B}$ for $M$ and $RMS$ did not decay at all over epochs ($E_{B}(M(\mathrm{epoch})) = -0.0848\frac{\Delta}{\mathrm{epoch}}, -0.1461\frac{\Delta}{\mathrm{epoch}}$, and $-0.0935\frac{\Delta}{\mathrm{epoch}}$; $E_{B}(RMS(\mathrm{epoch})) = -0.0846\frac{\Delta}{\mathrm{epoch}}, -0.1480\frac{\Delta}{\mathrm{epoch}}$, and $-0.0937\frac{\Delta}{\mathrm{epoch}}$ for healthy controls, CHF nonsurvivors, and CHF survivors, respectively). $E_{B}$ remained unchanged over a progressively larger number of epochs for all three groups (\textit{colored lines} in \textbf{Figs.~\ref{fig: HRV-2}a, c}). In contrast, $E_{B}$ for $M$ and $RMS$ of the shuffled RRi series rapidly decayed to a finite asymptotic value ($E_{B}(M(\mathrm{epoch})) = -1.2337\frac{\Delta}{\mathrm{epoch}}, -1.2890\frac{\Delta}{\mathrm{epoch}}$, and $-1.2275\frac{\Delta}{\mathrm{epoch}}$; $E_{B}(RMS(\mathrm{epoch})) = -1.2400\frac{\Delta}{\mathrm{epoch}}, -1.2951\frac{\Delta}{\mathrm{epoch}}$, and $-1.2321\frac{\Delta}{\mathrm{epoch}}$ for healthy controls, CHF nonsurvivors, and CHF survivors, respectively; \textit{grey lines} in \textbf{Figs.~\ref{fig: HRV-2}a, c}). In other words, $M$ and $RMS$-based HRV parameters failed to provide ergodic descriptions of HRV. Furthermore, the contrast between behaviors of $E_{B}$ for the original and the shuffled RRi series indicates that the very temporal structure of HRV contributes to this failure.

To test the specificity and reliability of these HRV parameters, we also performed Monte Carlo simulations by randomly sampling the $1000$-sample RRi series from the $24$-hour recordings for each individual and performing one-way ANOVA tests separately on these series' $M$ and $RMS$ values. We repeated this process $1000$ times. One-way ANOVAs failed to detect reduced $M$ of HRV due to the CHF, $36.8\%$ and $12.7\%$ times in nonsurvivors (\textit{red histogram} in \textbf{Fig.~\ref{fig: HRV-2}b}) and survivors (\textit{green histogram} in \textbf{Fig.~\ref{fig: HRV-2}b}), respectively, compared to healthy controls. Likewise, one-way ANOVAs failed to detect reduced $RMS$ of HRV due to the CHF, $34.6\%$ and $11.3\%$ times in nonsurvivors (\textit{red histogram} in \textbf{Fig.~\ref{fig: HRV-2}d}) and survivors (\textit{green histogram} in \textbf{Fig.~\ref{fig: HRV-2}d}), respectively, compared to healthy controls. In other words, we found a high likelihood of failing to identify statistically significant differences among the three groups' $M$ and $RMS$. These results confirm that linear descriptors $M$ and $RMS$ cannot be used as reliable HRV parameters for digital biomarkers of health and disease.

\subsection*{Linear descriptors $\mathrm{NN50}$ and $\mathrm{pNN50}$ are only weakly ergodic but not specific}

\begin{figure}[bt!]
\centering
\includegraphics[width=7in]{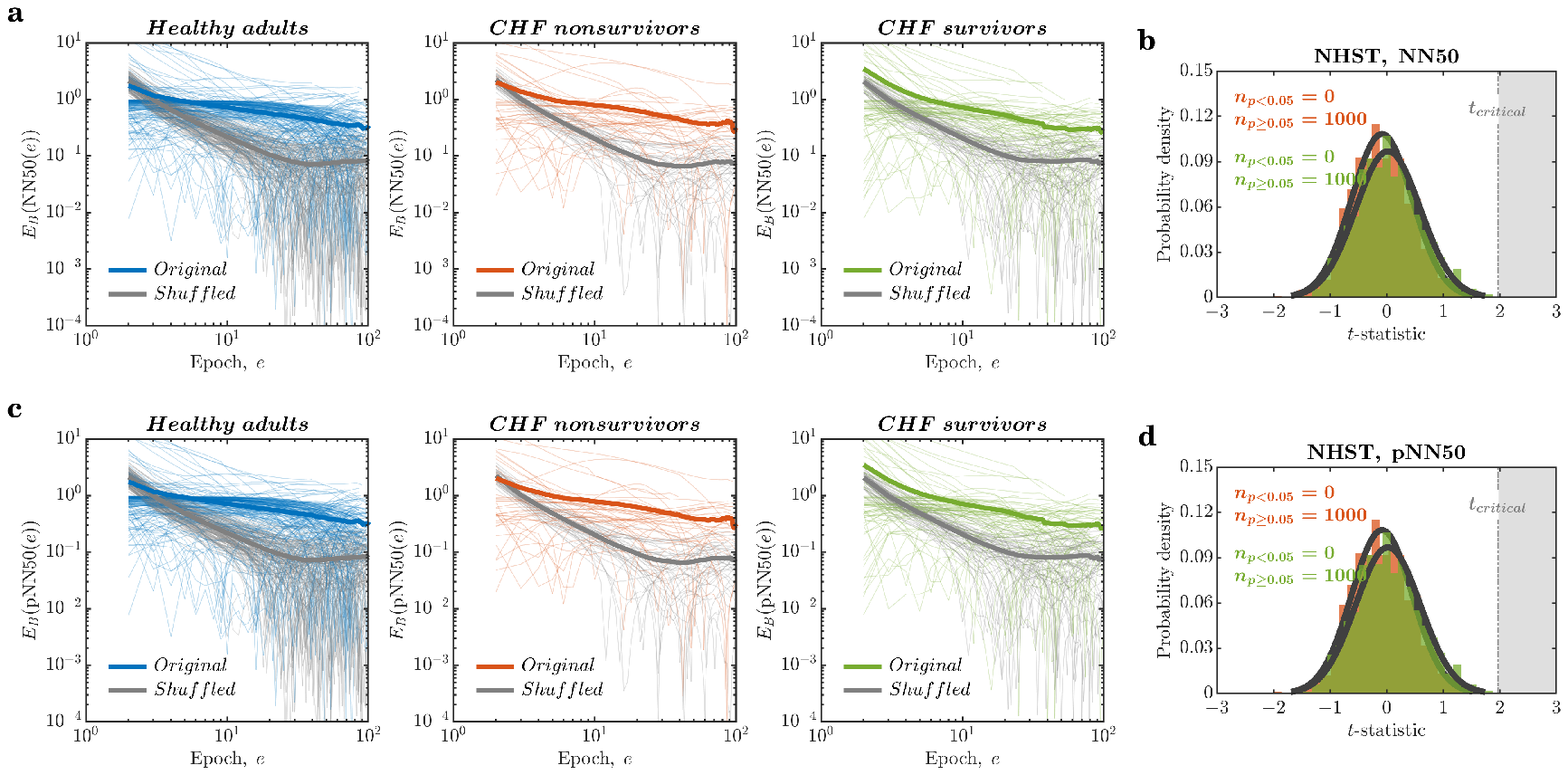}
\caption{\textbf{$\mathrm{NN50}$ and $\mathrm{pNN50}$ are specific but only weakly ergodic.} \textbf{a.} The epoch series of $\mathrm{NN50}$ describing the original RRi series show an initial decay in the ergodicity breaking parameter, $E_{B}$, with epochs (\textit{colored lines}), albeit shallower compared to the epoch series of $\mathrm{NN50}$ describing the shuffled RRi series (\textit{grey lines}). \textbf{b.} Null hypothesis significance testing (NHST) for $\mathrm{NN50}$ across the three groups. One-way ANOVAs revealed that $NN50$ of HRV did not differ between either patient populations and healthy controls: CHF nonsurvivors and survivors (\textit{red histogram} and \textit{green histogram}, respectively). \textbf{c.} The epoch series of $\mathrm{pNN50}$ describing the original RRi series show an initial decay in $E_{B}$ over epochs (\textit{colored lines}), albeit shallower compared to the epoch series of $\mathrm{pNN50}$ describing the shuffled RRi series (\textit{grey lines}). \textbf{d.} NHST for $\mathrm{pNN50}$ across the three groups. One-way ANOVAs revealed that $pNN50$ of HRV did not differ between either patient populations and healthy controls: CHF nonsurvivors and survivors (\textit{red histogram} and \textit{green histogram}, respectively). \textit{Thin lines} and \textit{thick lines} in \textbf{a, c} represent $E_{B}$ for individuals and mean $E_{B}$ for the three groups, respectively.}
\label{fig: HRV-3}
\end{figure}

The number of adjacent RR intervals that differ by more than $50$ milliseconds and the percentage of such RR intervals are two other linear descriptors widely used as HRV parameters \cite{shaffer2017overview}. $E_{B}$ for $\mathrm{NN50}$ and $\mathrm{pNN50}$ had similar behavior to that of the shuffled RRi series; however, $E_{B}$ had a shallower initial decay for $\mathrm{NN50}$ and $\mathrm{pNN50}$ with a progressively larger number of epochs ($E_{B}(\mathrm{NN50}(\mathrm{epoch})) = -0.4297\frac{\Delta}{\mathrm{epoch}}, -0.4281\frac{\Delta}{\mathrm{epoch}}$, and $-0.7090\frac{\Delta}{\mathrm{epoch}}$; $E_{B}(\mathrm{NN50}(\mathrm{epoch})) = -0.43\frac{\Delta}{\mathrm{epoch}}, -0.43\frac{\Delta}{\mathrm{epoch}}$, and $-0.71\frac{\Delta}{\mathrm{epoch}}$ for healthy controls, CHF nonsurvivors, and CHF survivors, respectively). Eventually, $E_{B}$ reached an asymptotic finite but a relatively larger value over a progressively larger number of epochs (\textit{colored lines} in \textbf{Fig.~\ref{fig: HRV-3}a, c}). These $E_{B}\mathrm{NN50}(\mathrm{epoch}))$ and $E_{B}(\mathrm{\mathrm{pNN50}} (\mathrm{epoch}))$ curves were only marginally shallower than those for the epoch series of $\mathrm{NN50}$ and $\mathrm{NN50}$ for the shuffled RRi series ($E_{B}(\mathrm{pNN50}(\mathrm{epoch})) = -1.2289\frac{\Delta}{\mathrm{epoch}}, -1.3169\frac{\Delta}{\mathrm{epoch}}$, and $-1.2557\frac{\Delta}{\mathrm{epoch}}$; $E_{B}(t_{MF}(\mathrm{epoch})) = -1.23\frac{\Delta}{\mathrm{epoch}}, -1.32\frac{\Delta}{\mathrm{epoch}}$, and $-1.26\frac{\Delta}{\mathrm{epoch}}$ for healthy controls, CHF nonsurvivors, and CHF survivors, respectively; \textit{grey lines} in \textbf{Fig.~\ref{fig: HRV-3}a, c}). Thus, $\mathrm{NN50}$ and $\mathrm{pNN50}$ break ergodicity only weakly, providing more ergodic descriptions of the non-ergodic HRV than the previous two linear descriptors, $M$ and $RMS$. Again, the contrast between the original and shuffled RRi series indicates that the very temporal structure of HRV contributes to this weak ergodicity breaking by $\mathrm{NN50}$ and $\mathrm{pNN50}$.

To test the specificity and reliability of these parameters, we performed Monte Carlo simulations by randomly sampling $1000$-sample RRi series from $24$-hour recordings for each individual and performing one-way ANOVA tests separately on these series’ $\mathrm{NN50}$ and $\mathrm{pNN50}$ values. We repeated this process $1000$ times. One-way ANOVAs revealed that $NN50$ of HRV did not differ between either patient populations and healthy controls: CHF nonsurvivors and survivors (\textit{red histogram} and \textit{green histogram}, respectively, in \textbf{Fig.~\ref{fig: HRV-3}b}). Likewise, one-way ANOVAs revealed that $pNN50$ of HRV did not differ between either patient populations and healthy controls: CHF nonsurvivors and survivors (\textit{red histogram} and \textit{green histogram}, respectively, in \textbf{Fig.~\ref{fig: HRV-3}b}). Hence, $\mathrm{NN50}$ and $\mathrm{pNN50}$ might only weakly break ergodicity but also not diagnose CHF.

\subsection*{Cascade-dynamical descriptors $H_{fGn}$ and $t_{MF}$ are both ergodic and specific}

We hypothesized that cascade-dynamical descriptors might provide ergodic descriptions of the non-ergodic HRV by capturing the source of ergodicity breaking. The most compelling descriptions of cascading dynamics come from multifractal geometry \cite{ihlen2010interaction,lovejoy2018weather,kelty2013tutorial}. Simulations of cascade processes show two critical features: long-range linear temporal correlations and nonlinear correlations involving interactions across timescales. The former feature appears most frequently as a fractional Gaussian noise (fGn) in which the standard deviation increases as a power function of the timescale. The fractional power in this function is known as the Hurst exponent, $H_{fGn}$. $H_{fGn}$ and has already been shown to be sensitive to differences in HRV due to congestive heart failure \cite{peng1995fractal}. The latter feature of nonlinear correlations concerns the effects spreading across the hierarchical organization of biological structures producing a variation in $H_{fGn}$, i.e., multifractality. We can estimate this nonlinear variation by estimating the variation in power functions over time and then comparing this multifractal variation to what a linear model of the underlying RRi series can produce. That is, by comparing the multifractality (i.e., the number of power functions) estimable for the original RRi series to the same multifractal property for a sample of synthetic RRi series. The one-sample $t$-test comparing the multifractality of the original to the synthetic RRi series provides a $t$-statistic, multifractal nonlinearity, $t_{MF}$, which quantifies nonlinear correlations due to cascade dynamics \cite{kelty2022multifractal}. $H_{fGn}$ and $t_{MF}$ have been shown to provide ergodic descriptions of the non-ergodic series of both simulated and empirical biological measurements \cite{kelty2022fractal,kelty2023multifractal,mangalam2022ergodic}. We aim to determine whether $H_{fGn}$ and $t_{MF}$ can adequately describe the non-ergodic HRV. Furthermore, we test whether $H_{fGn}$ and $t_{MF}$ provide specificity.

\begin{figure}[bt!]
\centering
\includegraphics[width=7in]{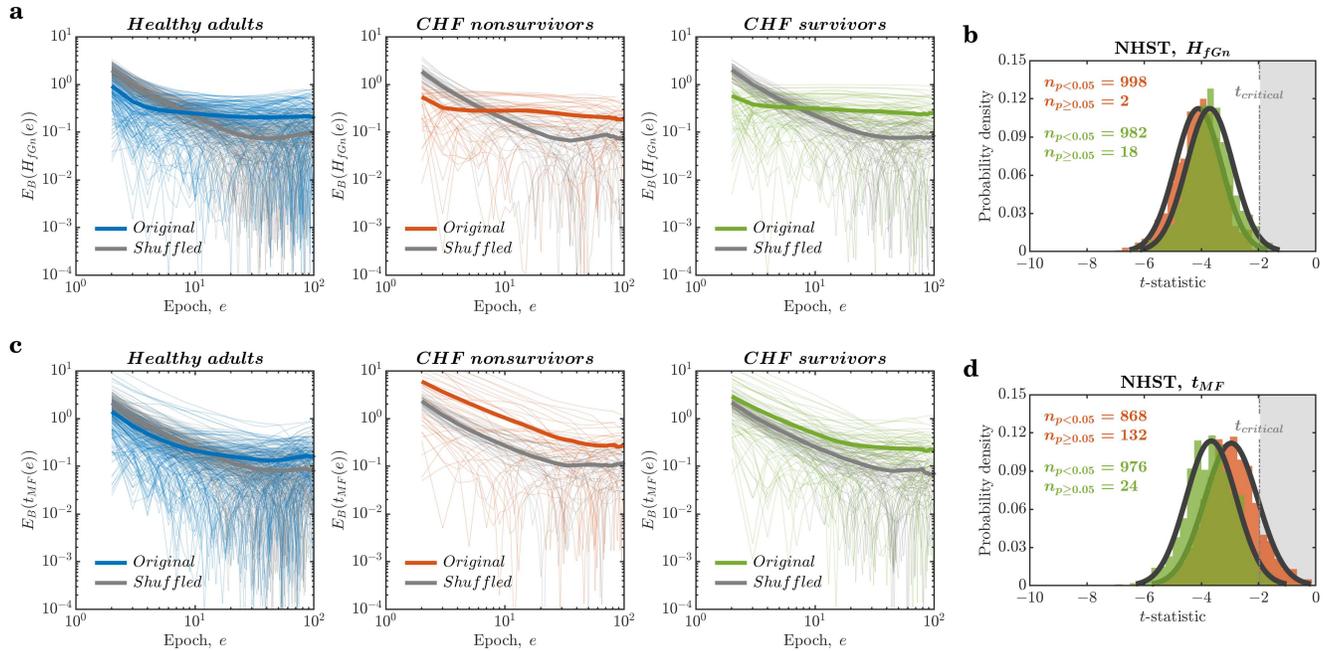}
\caption{\textbf{Cascade-dynamical descriptors of long-range correlations, $H_{fGn}$, and multifractal nonlinearity, $t_{MF}$, are ergodic and specific} \textbf{a.} The ergodicity breaking parameter, $E_{B}$, decayed initially for the $H_{fGn}$ of the original RRi series over epochs (\textit{colored lines}). However, this decay was shallower compared to that of the $H_{fGn}$ of the shuffled RRi series (\textit{grey lines}). \textbf{b.} NHST of $H_{fGn}$ across the three groups. One-way ANOVAs failed to detect reduced $H_{fGn}$ of HRV due to the CHF, only $0.2\%$ and $1.8\%$ times in nonsurvivors (\textit{red histogram}) and survivors (\textit{green histogram}), respectively, compared to healthy controls. \textbf{c.} $E_{B}$ decayed rapidly over epochs for the $t_{MF}$ of both the original RRi series (\textit{colored lines}) and the shuffled RRi series (\textit{grey lines}). \textbf{d.} One-way ANOVAs failed to detect reduced $t_{MF}$ of HRV due to the CHF, $13.2\%$ and $2.4\%$ times in nonsurvivors (\textit{red histogram}) and survivors (\textit{green histogram}), respectively, compared to healthy controls. \textit{Thin lines} and \textit{thick lines} in \textbf{a, c} represent $E_{B}$ for individuals and mean $E_{B}$ for the three groups, respectively.}
\label{fig: HRV-4}
\end{figure}

The behavior of $E_{B}$ for the epoch series of $H_{fGn}$ and $t_{MF}$ bears a strong resemblance to the behavior of $E_{B}$ for the shuffled RRi series. For $H_{fGn}$, $E_{B}$ had an initial shallower decay over a progressively larger number of epochs ($E_{B}(\Delta \alpha(\mathrm{epoch})) = -0.4640\frac{\Delta}{\mathrm{epoch}}, -0.1648\frac{\Delta}{\mathrm{epoch}}$, and $-0.2121\frac{\Delta}{\mathrm{epoch}}$; \textit{colored lines} in \textbf{Fig.~\ref{fig: HRV-4}a}). For $t_{MF}$, $E_{B}$ rapidly decayed initially over a progressively larger number of epochs ($E_{B}(t_{MF}(\mathrm{epoch})) = -0.8372\frac{\Delta}{\mathrm{epoch}}, -1.0238\frac{\Delta}{\mathrm{epoch}}$, and $-1.1759\frac{\Delta}{\mathrm{epoch}}$; \textit{colored lines} in \textbf{Fig.~\ref{fig: HRV-4}c}). These $E_{B}(H_{fGn}(\mathrm{epoch}))$ curves show a faster decay for the shuffled RRi series ($E_{B}(H_{fGn}(\mathrm{epoch})) = -1.2423\frac{\Delta}{\mathrm{epoch}}, -1.2434\frac{\Delta}{\mathrm{epoch}}$, and $-1.2002\frac{\Delta}{\mathrm{epoch}}$; \textit{grey lines} in \textbf{Fig.~\ref{fig: HRV-4}a}). However, the decay rate of $E_{B}(t_{MF}(\mathrm{epoch}))$ curves for the shuffled RRi series was comparable to that of the original RRi series ($E_{B}(t_{MF}(\mathrm{epoch})) = -1.2091\frac{\Delta}{\mathrm{epoch}}, -1.1555\frac{\Delta}{\mathrm{epoch}}$, and $-0.9862\frac{\Delta}{\mathrm{epoch}}$ for healthy controls, CHF nonsurvivors, and CHF survivors, respectively; \textit{grey lines} in \textbf{Fig.~\ref{fig: HRV-4}c}). These results show that the cascade-dynamical descriptors, $H_{fGn}$ and $t_{MF}$, provide ergodic descriptions of the non-ergodic HRV. The cascade-dynamical nature of HRV that contributed to the non-ergodicity of linear descriptors like $M$ and $RMS$ was appropriately captured by cascade-dynamical descriptors $H_{fGn}$ and $t_{MF}$.

To test the specificity of these cascade-dynamical descriptors, we performed Monte Carlo simulations by randomly sampling $1000$-sample RRi series from $24$-hour recordings for each individual and performing one-way ANOVA tests separately on these series’ $H_{fGn}$ and $t_{MF}$ values. We repeated this process $1000$ times. One-way ANOVAs failed to detect reduced $H_{fGn}$ of HRV due to the CHF, $0.2\%$ and $1.8\%$ times in nonsurvivors (\textit{red histogram} in \textbf{Fig.~\ref{fig: HRV-4}b}) and survivors (\textit{green histogram} in \textbf{Fig.~\ref{fig: HRV-4}b}), respectively, compared to healthy controls. Likewise, one-way ANOVAs failed to detect reduced $M$ of HRV due to the CHF, $13.2\%$ and $2.4\%$ times in nonsurvivors (\textit{red histogram} in \textbf{Fig.~\ref{fig: HRV-4}b}) and survivors (\textit{green histogram} in \textbf{Fig.~\ref{fig: HRV-4}b}), respectively, compared to healthy controls. Thus, $H_{fGn}$ and $t_{MF}$ provide ergodic descriptions of the non-ergodic HRV and can specifically differentiate clinical groups with high reliability. These results support our proposal of capitalizing cascade-dynamical descriptors as generalizable and reproducible HRV parameters for digital biomarkers of health and disease.

\begin{figure}[bt!]
\centering
\includegraphics[width=5in]{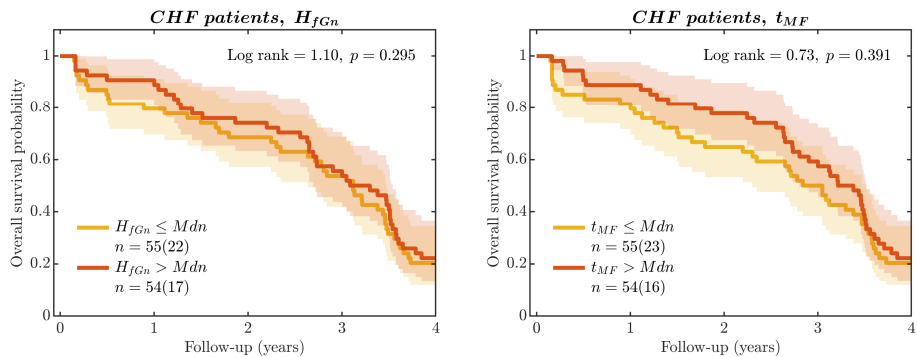}
\caption{\textbf{Kaplan-Meier cumulative survival curves of patients with congestive heart failure---both nonsurvivors and survivors.} \textbf{a.} Stratified to patients with long-range correlations in HRV, $H_{fGn}\leq Mdn$ and $H_{fGn}>Mdn$, with log-rank statistics. \textbf{b.} Stratified to patients with multifractal nonlinearity in HRV, $t_{MF}\leq Mdn$ and $t_{MF}>Mdn$, with log-rank statistics. The cutoff points for dichotomization were determined by the respective descriptor's median ($Mdn$). \textit{Shaded areas} indicate $95\%$ confidence intervals. $n$ denotes the number of patients in a subgroup, with the number of deaths during the observation period in \textit{parentheses}. The obtained hazard ratios $0.814 [0.619,1.010], p=0.295$ for $H_{fGn}$ and $0.846 [0.652,1.040],p=0.391$ for $t_{MF}$ indicate that these descriptors are not sufficient for reliable prognostics.}
\label{fig: HRV-5}
\end{figure}

To also investigate the prognostic capacities of the cascade-dynamical descriptors $H_{fGn}$ and $t_{MF}$, we performed survival analysis to examine if these descriptors could predict mortality among CHF patients \cite{efron1988logistic,rich2010practical}. \textbf{Fig.~\ref{fig: HRV-5}} shows Kaplan-Meier cumulative survival curves using $H_{fGn}$ and $t_{MF}$ as predictors. The two descriptors' median ($Mdn$) determined the cutoff points for dichotomization. $H_{fGn}$ and $t_{MF}$ failed to predict mortality and produced comparable Mantel-Haenszel log-rank statistics. These hazard ratios were obtained: $0.814 [0.619,1.010], p=0.295$ for $H_{fGn}$ and $0.846 [0.652,1.040],p=0.391$ for $t_{MF}$.

\section*{Discussion}

Here, we surveyed various descriptors that could be used by traditional and digital medicine to inform the diagnosis and prognosis of cardiovascular conditions such as CHF to identify those descriptors that could provide predictive, specific, generalizable, and reproducible assessments. The primary risk we have highlighted is that the raw RRi series breaks ergodicity. This non-ergodicity of HRV is a liability to clinical care because the raw RRi series fail to converge toward an average. Without this convergence, any sequence of RRi cannot be deemed sufficiently representative---whether of the patient's long-term HRV or groups with a definitive clinical diagnosis or prognosis. We found that many of the most conventional linear descriptors are all as non-ergodic as the raw RRi series they summarize. We also identified that the primary source of these non-ergodicities is the very temporal structure of HRV and its cascade-dynamical nature. Afterward, we hypothesized that this very origin of non-ergodicities might hold the key to the ergodic descriptions of HRV. The cascade-dynamical descriptors $H_{fGn}$ and $t_{MF}$ confirmed this hypothesis, provided ergodic descriptions of the non-ergodic HRV, and could also specifically differentiate clinical groups. However, our survival analysis indicated that even $H_{fGn}$ and $t_{MF}$ cannot sufficiently predict post-CHF prognosis. Previous works exploring the usefulness of cascade-dynamical descriptors have often relied on a suite of descriptors, including those sensitive to the non-Gaussian statistics of HRV \cite{hayano2011increased,kiyono2004critical,kiyono2005multiscale,kiyono2005phase,kiyono2007estimator,kiyono2008non,kiyono2009log,kiyono2012non}. This suggests that it might be best to employ cascade-dynamical descriptors, like $H_{fGn}$ and $t_{MF}$, alongside more traditional descriptors to achieve maximum predictivity, specificity, generalizability, and reproducibility.

The widespread adoption of smartwatches and other wearable biosensors with heart-rate monitoring capabilities has sparked hope for the early detection of cardiovascular diseases. However, the belief that more data is sufficient to improve predictions is overly optimistic and misguided. Machine learning/artificial intelligence (ML/AI) models are being developed to achieve highly accurate, sensitive, and specific measures of cardiovascular health; however, to aptly capitalize on these powerful tools, the need for a theoretical understanding of heart rate variability must be addressed. Despite the availability of vast amounts of data, the advancement in understanding the nature of heart rate variability has been modest despite years of work and thousands of scientific publications. This limitation prevents us from making meaningful inferences using ML/AI models. The current reliance on manual or automatic feature extraction \cite{coutts2020deep,hannun2019cardiologist,hong2020opportunities,kim2016data,torres2020multi,wang2019detection} is problematic since these features may not suitably reflect the primary causal mechanisms and be too much dependent on contextual variables \cite{baek2017effect,faust2020deviations,kim2009effect,kim2012effect}. This study emphasizes that the current optimism surrounding the use of wearables and ML/AI models to detect cardiovascular diseases must be accompanied by a deeper understanding of the ergodicity-breaking behavior of HRV.

Our conclusions merit urgent attention as they show the unreliability of prevalent linear descriptors of HRV like mean-based parameters. We have shown that some of the most intuitive conventional descriptors of HRV, like mean-based descriptors, are non-ergodic. In contrast, cascade-dynamical descriptors, such as $t_{MF}$, can significantly improve the assessment of cardiovascular health. These results align with those of previous studies that had reported that nonlinear descriptors could provide additional prognostic information compared to conventional linear descriptors \cite{perkiomaki2011heart,voss2009methods,kiyono2008non,hayano2011increased,kiyono2012non}, e.g., short-term scaling exponent is a better predictor of mortality or other primary endpoints in cardiovascular patients \cite{makikallio1999fractal,huikuri2000fractal}. Moreover, such nonlinear descriptors have even been found to be reproducible across different populations \cite{maestri2007assessing,perkiomaki2001fractal,pikkujamsa2001determinants} and contexts, e.g., receiving or not receiving beta-blockers \cite{jokinen2003temporal}, and different times and methods of measurement \cite{perkiomaki2001fractal,perkiomaki2011heart}.

Our results also show that a descriptor's ergodicity is necessary but insufficient for its prognostic capability. Although $H_{fGn}$ and $t_{MF}$ provided ergodic descriptions of HRV, they failed to predict mortality in CHF patients. Thus, although ergodicity is necessary for generalizable and reproducible inferences, to reach the utmost specific, generalizable, and reproducible assessment, combining descriptors that provide ergodic descriptions, like the cascade-dynamical descriptors investigated here, with other descriptors, like the conventional ones.

To also compare the nonlinear descriptors investigated here, it must be noted that $H_{fGn}$ is primarily a monofractal descriptor and is best suited to describe series generated based on one fractal-scaling exponent. However, the modeling of cascade dynamics due to nonlinear interactions across scales inherent to HRV is beyond the scope of $H_{fGn}$ and requires multifractal formalism \cite{ihlen2010interaction,lovejoy2018weather,mandelbrot1974intermittent}. Monofractal fluctuations such as fGn are ideally defined exhaustively by single fractional exponents $H_{fGn}$ and fall cleanly within the linear model through an autocorrelation function indicative of fractional integration \cite{mandelbrot1968fractional,mandelbrot1982fractal}. The nonlinearity of interactions across scales requires not only one but many fractional scaling exponents in addition to strictly linear long-range correlations. Hence, multifractal modeling is necessary to analyze the putative cascade-dynamical route to non-ergodicity thoroughly \cite{kelty2022fractal,kelty2023multifractal,mangalam2022ergodic}, i.e., the inherently multifractal descriptor, $t_{MF}$, is superior to $H_{fGn}$ in encoding nonlinear interactions across scales, which is characteristic of HRV.

Some other points also warrant further attention. For more comprehensive employment of nonlinear descriptors, such as those proposed here, especially in traditional medical settings, it might be necessary to provide more intuitive interpretations for clinicians and educate clinicians so that the biological basis of these mathematical parameters is clear \cite{captur2017fractal}. Also, cascade dynamics is only one of the mechanisms that can lead to ergodicity-breaking physiological variabilities. It is still being determined whether all such mechanisms could be modeled as cascade processes (e.g., \cite{mangalam2023ergodic}). Despite the central role of cascade processes in biomedical explanation \cite{partridge2010contemporary}, we hope that future investigations examine a broader class of anomalous diffusion regimes \cite{barkai2012single,fulinski2011anomalous,krapf2019strange,magdziarz2011anomalous,metzler2012role,metzler2014anomalous,munoz2021objective,thiel2014weak,vinod2022nonergodicity,wang2022restoring} that can also lead to ergodicity-breaking physiological variabilities. Further work is needed to determine whether cascade-dynamical descriptors enable reproducible health assessment when the sources of ergodicity breaking are more nuanced. The statistical modeling framework presented in the present study will be fundamental in guiding these investigations.

Eventually, the challenges faced in this study and our proposed solutions should be unrestricted to the case of HRV and cardiovascular health. Non-ergodicity and cascade dynamics abound in biological processes and are regularities---not exceptions. Much more attention must be paid to the ergodicity of investigated biological phenomena. Moreover, in cases of ergodicity breaking, we have shown here and in previous studies \cite{kelty2022fractal,kelty2023multifractal,mangalam2022ergodic} that cascade dynamics should be considered one of the primary candidates for its origin and that capturing this origin through nonlinear, multifractal, and cascade-dynamical descriptors may be the key to ergodically describing non-ergodic phenomena. The importance of these insights cannot be exaggerated as they are crucial for reliable and reproducible diagnosis and prognosis across all fields. Non-ergodicity may be a signature of life, but seeking ergodicity in our generalizations and causal reasoning is pivotal for arriving at generalizable and reproducible digital biomarkers of health and disease.

\section*{Methods}

Each patient gave informed written consent with full knowledge of the details. The ethics committee of Fujita Health University approved the research, which followed the guidelines stated in the Declaration of Helsinki. All data were fully anonymized before we accessed them.

\subsection*{Subjects}

Based on the data of one of our previous studies \cite{kiyono2008non}, we retrospectively enrolled the patients referred to the hospital of the Fujita Health University from January 2000 to December 2001 for assessment or treatment of CHF. $24$-hour monitoring of Holter ECG was conducted before their hospital discharge. To be eligible for this study, the patients had to be in normal sinus rhythm and had Holter ECG recordings whose periods taken up by artifacts, or noise were less than $5\%$. No intravenous positive-inotropic agents or vasodilators were administered during the Holter ECG recordings. We excluded patients with chronic or paroxysmal atrial fibrillation, permanent or temporary cardiac pacemakers, active thyroid disease, or malignancy.

\subsection*{Follow-up and endpoint}

We recorded the baseline data upon hospital discharge and the time-to-event information for each subject in a database. We then periodically sent questionnaires to patients or their families during the follow-up period and conducted telephone interviews to gather mortality information. \textit{Death} from progressive heart failure was defined as death resulting from multi-organ failure caused by the progression of pump failure, and sudden death was defined as either witnessed cardiac arrest or death within one hour of onset of acute symptoms or the unexpected death of a patient known to have been well within the previous 24 hours.

\subsection*{Analysis of holter ECG}

We digitized ECG signals at 125 Hz and 12 bits using proprietary software (Cardy Analyzer II, Suzuken Co., Ltd., Nagoya, Japan) and included only recordings with at least 22 hours of data in the analysis and $>95\%$ of quantified sinus beats. Although the Cardy Analyzer II software had detected and labeled all QRS complexes in each recording, we manually corrected any errors in R-wave detection and QRS labeling. We then exported the individual files containing the duration of individual RRi intervals and morphology classifications of individual QRS complexes (normal, supraventricular, and ventricular premature complexes, supraventricular, and ventricular escape beats). We analyzed the $24$-hour sequence of intervals between two successive R waves of sinus rhythm (i.e., heart rate variability or HRV). To avoid the adverse effects of any remaining errors in detecting the R wave, we reviewed large ($>20\%$) consecutive RRi interval differences until all errors were corrected. In addition, when we encountered atrial or ventricular premature complexes, we interpolated the corresponding RRi intervals by the median of the two successive beat-to-beat intervals. We also confirmed that no sustained tachyarrhythmias were present in the HRV recordings. We then interpolated the observed RRi series with a cubic spline function and resampled at an interval ($\Delta t$) of $500$ ms ($2$ Hz), yielding interpolated RRi series.

\subsection*{Estimating descriptors of HRV for epoch series}

We computed the following descriptors of HRV---linear descriptors over non-overlapping $500$-beat epochs extracted from the RRi series and fractal and multifractal descriptors over non-overlapping $1000$-sample epochs extracted from the interpolated RRi series. Hence, we computed fractal and multifractal descriptors in the time domain as both these are time-domain analytical methods. We computed these descriptors for the original (i.e., unshuffled) and a shuffled counterpart (i.e., a version with the temporal information destroyed) of each RRi series.

\subsubsection*{Conventional linear descriptors}

We computed four linear descriptors of HRV. (i) Mean of successive RR intervals ($M$). (ii) Root mean square of successive RR intervals ($RMS$) mathematically defined as
\begin{equation*}
  RMS = \sqrt{\frac{1}{T} \sum_{t=1}^{T} |x(t)|^{2}}. \tag{6}\label{eq:6}
\end{equation*}
(iii) Number of pairs of successive RRi intervals that differ by more than $50$ ms ($\mathrm{NN50}$). (iv) The percentage of successive RRi intervals that differ from each other by more than $50$ ms ($\mathrm{pNN50}$).

\subsubsection*{Fractal-scaling descriptor of long-range correlations using monofractal detrended fluctuation analysis}

Detrended fluctuation analysis (DFA) computes the Hurst exponent, $H_{fGn}$, quantifying the strength of long-range correlations in series \cite{peng1994mosaic,peng1995quantification} using the first-order integration of $T$-length series $x(t)$:
\begin{equation*}
  y(i) = \sum_{k=1}^{i} \Bigl ( x(k) - \overline{x(t)} \Bigl ), \quad i = 1,2,3, \dots, T. \tag{7}\label{eq:7}
\end{equation*}
DFA computes root mean square ($RMS$; i.e., averaging the residuals) for each linear trend $y_{n}(t)$ fit to $N_n$ non-overlapping $n$-length bins to build a fluctuation function: 
\begin{equation*}
  f(v, n) = \sqrt{\frac{1}{N_n} \sum_{v=1}^{N_n} \biggl (\frac{1}{n} \sum_{i=1}^{n} \Bigl ( y \bigl ( (v-1)\,n+i \bigl ) - y_{v}(i) \Bigl ) ^{2} \biggl ) }, \quad n = \{4, 8, 12, \dots \} < T/4. \tag{8}\label{eq:8}
\end{equation*}
$f(n)$ is a power law,
\begin{equation*}
  f(n) \sim n^{H_{fGn}}, \tag{9}\label{eq:9}
\end{equation*}
where $H_{fGn}$ is the scaling exponent estimable using logarithmic transformation:
\begin{equation*}
  \log f(n) = H_{fGn}\log n. \tag{10}\label{eq:10}
\end{equation*}
Higher $H_{fGn}$ corresponds to stronger long-range correlations.

\subsubsection*{Multifractal spectrum width based on the direct estimation of singularity spectrum}

Chhabra and Jensen’s \cite{chhabra1989direct} direct method estimates multifractal spectrum width $\Delta\alpha$ by sampling a series $x(t)$ at progressively larger scales using the proportion of signal $P_{i}(n)$ falling within the $v$th bin of scale $n$ as
\begin{equation*}
  P_{v}(n) = \frac{\sum\limits_{k = (v-1)\,n+1}^{N_{n}} x(k)}{\sum{x(t)}}, \quad n = \{2, 4, 8, 16, \dots \} < T/8. \tag{11}\label{eq:11}
\end{equation*}
As $n$ increases, $P_{v}(n)$ represents a progressively larger proportion of $x(t)$,
\begin{equation*}
  P(n) \propto n^{\alpha}, \tag{12}\label{eq:12}
\end{equation*}
suggesting a growth of the proportion according to one ``singularity'' strength $\alpha$ \cite{mandelbrot1982fractal}. $P(n)$ exhibits multifractal dynamics when it grows heterogeneously across time scales $n$ according to multiple singularity strengths, such that
\begin{equation*}
  P(n_{v}) \propto n^{\alpha_{v}}, \tag{13}\label{eq:13}
\end{equation*}
whereby each $v$th bin may show a distinct relationship of $P(n)$ with $n$. The width of this singularity spectrum, $\Delta\alpha = (\alpha_{max}-\alpha_{min})$, indicates the heterogeneity of these relationships \cite{halsey1986fractal,mandelbrot2013fractals}.

Chhabra and Jensen's \cite{chhabra1989direct} method estimates $P(n)$ for $N_{n}$ non-overlapping bins of $n$-sizes and transforms them into a ``mass'' $\mu(q)$ using a $q$ parameter emphasizing higher or lower $P(n)$ for $q>1$ and $q<1$, respectively, in the form
\begin{equation*}
  \mu_{v}(q,n) = \frac{ \bigl [P_{v}(n) \bigl ] ^{q}}{\sum\limits_{j=1}^{N_{n}} \bigl[ P_{j}(n) \bigl ] ^{q}}. \tag{14}\label{eq:14}
\end{equation*}
Then, $\alpha(q)$ is the singularity for mass $\mu$-weighted $P(n)$ estimated as
\begin{equation*}
  \alpha(q) = - \lim_{N_{n}\to\infty} \frac{1}{\ln N_{n}} \sum_{v=1}^{N_{n}} \mu_{v} (q,n) \ln P_{v}(n)
\end{equation*}
\begin{equation*}
  = \lim_{n\to0} \frac{1}{\ln n} \sum_{v=1}^{N_{n}} \mu_{v} (q,n) \ln P_{v}(n). \tag{15}\label{eq:15}
\end{equation*}
Each estimated value of $\alpha(q)$ belongs to the multifractal spectrum only when the Shannon entropy of $\mu(q,n)$ scales with $n$ according to the Hausdorff dimension $f(q)$ \cite{chhabra1989direct}, where
\begin{equation*}
  f(q) = - \lim_{N_{n}\to\infty} \frac{1}{\ln N_{n}} \sum_{v=1}^{N_{n}} \mu_{v} (q,n) \ln \mu_{v}(q,n)
\end{equation*}
\begin{equation*}
  = \lim_{v\to0} \frac{1}{\ln n} \sum_{v=1}^{N_{n}} \mu_{v} (q,n) \ln \mu_{v}(q,n). \tag{16}\label{eq:16}
\end{equation*}

For values of $q$ yielding a strong relationship between Eqs.~(\ref{eq:15}) \& (\ref{eq:16})---in this study, correlation coefficient $r > 0.9975$, the parametric curve $(\alpha(q),f(q))$ or $(\alpha,f(\alpha))$ constitutes the multifractal spectrum and $\Delta \alpha$ (i.e., $\alpha_{max}-\alpha_{min}$) constitutes the multifractal spectrum width. $r$ determines that only scaling relationships of comparable strength can support the estimation of the multifractal spectrum, whether generated as cascades or surrogates. Using a correlation benchmark aims to operationalize previously raised concerns about mis-specifications of the multifractal spectrum \cite{zamir2003critique}.

\subsubsection*{Surrogate testing using Iterated Amplitude Adjusted Fourier Transformation (IAAFT) generated \textit{t}-statistic, $t_{MF}$}

While multifractality is necessary for cascade-like interactivity, multifractality is not conclusive evidence of cascade-like interactivity, as it can follow from other sources, e.g., linear autocorrelation and outliers in the histogram \cite{veneziano1995multifractal}. To identify whether non-zero multifractal spectrum width (i.e., $\Delta \alpha > 0$) reflected multifractality due to nonlinear interactions across scales, we compared $\Delta \alpha$ for the original and shuffled RRi series to $\Delta \alpha$ for $32$ iterated amplitude adjusted Fourier transform (IAAFT) surrogates \cite{ihlen2012introduction,schreiber1996improved}. IAAFT randomizes original values time-symmetrically around the autoregressive structure, generating surrogates with randomized phase ordering of the series’ spectral amplitudes while preserving linear temporal correlations. We refer interesting readers to Kelty-Stephen et al. \cite{kelty2022multifractal} for a step-by-step guide to generating the IAAFT surrogates for any series. The resulting surrogate series should thus have the same values as the original series and thus the same mean and variance. It should also have the same amplitude spectrum and autocorrelation function as the original series. The one-sample $t$-statistic, $t_{MF}$ takes the subtractive difference between $\Delta \alpha$ for the original series and that for $32$ surrogates, dividing by the standard error of $\Delta \alpha$ for the surrogates.

\subsection*{Estimating ergodicity breaking parameter, $E_{B}$}

Ergodicity can be quantified using a dimensionless statistic of ergodicity breaking parameter, $E_{B}$, also known as the Thirumalai-Mountain metric \cite{he2008random,thirumalai1989ergodic} and already mentioned by Rytov et al. \cite{rytov1989principles}, computed as
\begin{equation*}
  E_{B}(x(t)) = \frac{ \Bigl \langle \Bigl [ \overline{\delta^{2}(x(t))} \Bigl ]^{2} \Bigl \rangle - \Bigl \langle \overline{\delta^{2}(x(t))} \Bigl \rangle^{2}}{ \Bigl \langle \overline{\delta^{2}(x(t))} \Bigl \rangle ^{2}}. \tag{17}\label{eq:17}
\end{equation*}
where $\overline{\delta^{2}(x(t))} = \int_{0}^{t - \Delta} [x(t^{\prime} + \Delta) - x(t^{\prime})]^{2}dt^{\prime} \bigl / (t - \Delta)$ is the time average mean-squared displacement of the stochastic series $x(t)$ for lag time $\Delta$. This relationship is effectively the variance of sample variance divided by the total-sample squared variance. Rapid decay of $E_{B}$ to a finite asymptotic value for progressively larger samples, i.e., $E_{B}\rightarrow0$ as $t \rightarrow \infty$ implies ergodicity. Thus, for Brownian motion $E_{B}(x(t)) = \frac{4}{3} (\frac{\Delta}{t})$ \cite{cherstvy2013anomalous,metzler2014anomalous}. Slower decay indicates less ergodic systems in which trajectories are less reproducible, and no decay or convergence to a finite asymptotic value indicates strong ergodicity breaking \cite{deng2009ergodic,wang2020fractional}. $E_{B}(x(t))$ thus allows testing whether a given series fulfills ergodic assumptions or breaks ergodicity. For instance, it has been shown that for fractional Brownian motion (FBM) \cite{deng2009ergodic,wang2020fractional},
\begin{equation*}
  E
_{B}(x(t)) =
  \begin{cases}
    k (H_{fGn}) \frac{\Delta}{t} & \text{if } 0 < H_{fGn} < \frac{3}{4} \\
    k (H_{fGn}) \frac{\Delta}{t}\ln{t} & \text{if } H_{fGn} = \frac{3}{4} \\
    k (H_{fGn}) ( \frac{\Delta}{t} ) ^{4 - 4H_{fGn}} & \text{if } \frac{3}{4} < H_{fGn} < 1.
  \end{cases}
  \tag{18}\label{eq:18}
\end{equation*}
The present work is less focused on firmly meeting the criterion of $E_{B}$ converging to zero within our finite samples. Instead, we compared the original and shuffled RRi series to assess ergodicity breaking instead of strict convergence of $E_{B}$ to zero. We computed $E_{B}$ for each original and shuffled RRi series (range = $T/50$; lag $\Delta$ = 10) and for each epoch series of $M$, $RMS$, $\mathrm{NN50}$, $\mathrm{PNN50}$, $H_{fGn}$, and $t_{MF}$ for the original and shuffled RRi series (range = $N_{\mathrm{epochs}}/2$; lag $\Delta$ = 1).

 \subsection*{Monte Carlo simulations}

We performed Monte Carlo simulations to test our hypothesis that ergodicity breaking by various linear and cascade-dynamical descriptors of HRV could compromise the reliability of these descriptors as diagnostic biomarkers. We randomly sampled $1000$-sample RRi series from $24$-hour recordings for each individual and performed linear mixed-effects models separately on $M$, $RMS$, $\mathrm{NN50}$, $\mathrm{pNN50}$, $H_{fGn}$, $t_{MF}$, values calculated from these series. We used linear mixed-effects models with each descriptor as the dependent variable and the participant group as the independent variable. The $t$-statistic and the resultant $p$ value were saved across the $1000$ iterations. We performed all mixed-effects modeling in MATLAB 2022b (Mathworks, Inc., Natick, MA) using the function \texttt{fitlme()}.

\subsection*{Survival analysis}

We examined whether $H_{fGn}$ and $t_{MF}$ were predictive of death using univariate Cox proportional hazards regression analysis \cite{efron1988logistic,rich2010practical}. We used the Mantel-Haenszel log-rank test to compare Kaplan-Meier cumulative survival curves to examine the impact of identified risk factors on survival. We performed all survival analysis in \texttt{R} \cite{team2013r} using the function \texttt{coxph()} from the package ``survival'' \cite{therneau2015package}.

\section*{Acknowledgements}

This work was supported by the Center for Research in Human Movement Variability at the University of Nebraska at Omaha, funded by the NIH award P20GM109090.

\section*{Author contributions statement}

\textbf{Madhur Mangalam:} Conceptualization, Methodology, Software, Validation, Formal analysis, Writing -- Original draft, Writing -- Review \& Editing, Visualization, Project administration; \textbf{Arash Sadri:} Writing -- Original draft, Writing -- Review \& Editing; \textbf{Junichiro Hayano:} Investigation, Data curation; \textbf{Eiichi Watanabe:} Investigation, Resources, Data curation; \textbf{Ken Kiyono:} Conceptualization, Methodology, Writing -- Review \& Editing; \textbf{Damian G. Kelty-Stephen:} Conceptualization, Methodology, Writing -- Original draft, Writing -- Review \& Editing.

\section*{Competing interests}

The authors declare no competing interests.

\section*{Data availability}

The data that support the results reported herein can be obtained upon request from the corresponding author.

\bibliography{sample}

\end{document}